\documentclass{ifacconf}
\usepackage{textcomp}
\usepackage{graphicx}      
\usepackage{natbib}        
\usepackage{amsfonts}
\usepackage{pgfplots}
\usepackage{rotating}
\pgfplotsset{width=8.4cm,compat=1.3}
\usepackage{enumerate}

\newtheorem{problem}{\bf Problem}[section]
\newtheorem{example}{\bf Example}[section]

\usepackage{algorithm}
\usepackage{algorithmicx,algpseudocode}

\usepackage{tikz}

\usepackage{tikz-qtree}
\usetikzlibrary{arrows}

\usepackage[cmex10]{amsmath}
\newcounter{row}
\newcounter{col}

\newcounter{row7}
\newcounter{col7}
\newcommand\setrowseven[7]{
  \setcounter{col7}{1}
  \foreach \n in {#1, #2, #3, #4, #5, #6, #7} {
    \edef\x{\value{col7} - 0.5}
    \edef\y{7.5 - \value{row7}}
    \node[anchor=center] at (\x, \y) {\n};
    \stepcounter{col7}
  }
  \stepcounter{row7}
}

\begin{document}
\begin{frontmatter}

\title{An Efficient Formula Synthesis Method with Past Signal Temporal Logic}

\author[First]{Mert Ergurtuna} 
\author[First]{Ebru Aydin Gol} 

\address[First]{Department of Computer Engineering, Middle East Technical University, Ankara Turkey \\(e-mail: mertergurtuna@gmail.com), (e-mail: ebru@ceng.metu.edu.tr) .}

\begin{abstract}                
In this work, we propose a novel method to find temporal properties that lead to the unexpected behaviors from labeled dataset.  We express these properties in past time Signal Temporal Logic (ptSTL). First, we present a novel approach for finding parameters of a template ptSTL formula, which extends the results on monotonicity based parameter synthesis. The proposed method optimizes a given monotone criteria while bounding an error. Then, we employ the parameter synthesis method in an iterative unguided formula synthesis framework. In particular, we combine optimized formulas iteratively to describe the causes of the labeled events while bounding the error. We illustrate the proposed framework on two examples.
\end{abstract}


\end{frontmatter}

\section{Introduction}

Designing cyber-physical systems that achieve complex tasks is a difficult and error prone process.  The resulting models are, in general, complex and composed of various sub-modules such as the Simulink models~\citep{MATLAB:20106}. Once the model is developed, it's traces are checked against the specifications for verification. While it is relatively easy to simulate the system and mark the unexpected behaviors, it is extremely challenging to locate the errors in the model that lead to the unexpected behaviors.
In this work, we propose a novel method to find temporal properties that lead to the unexpected behaviors from labeled system traces in an automated way. The properties generated by our approach can give an insight on the underlying cause and help the design engineer to identify the corresponding modeling errors. 

We express the temporal properties as signal temporal logic (STL) formulas. STL is a rich specification language that extends linear temporal logic~\citep{baierBook} and it is developed to describe properties of real valued signals~\citep{donze}. Due to its expressivity and efficient algorithms for checking continuous signals against STL formulas, it is used in different areas including runtime verification~\citep{Bartocci2018}, analysis of time series data~\citep{majumdar:2017} and formal control~\citep{Raman:2015}. 
An STL formula is defined using the Boolean and temporal operators, and predicates in the form of linear inequalities. In this work, we focus on the past time fragment of STL called ptSTL, where only the past time temporal operators are allowed. For example, $\mathbf{P}_{[0,3]} x > 5$ requires $x$ to exceed $5$ within the last $3$ seconds. More complex formulas can be easily written with combinations of Boolean and temporal operators. As an example, consider the following specification ``within the last 20 seconds, $x$ goes above $12$, and since then, within every five seconds, $y$ drops below $0$". The specification is expressed as ptSTL formula $\mathbf{P}_{[0,5]}\ y < 0\ \mathbf{S}_{[0,20]}\ x > 12$.

In this paper, we address the following problem: given a dataset of traces that are labeled at each time point (e.g. test results), find a ptSTL formula such that the evaluation of the formula along the traces will mimic the given labels. Essentially, the objective is to capture the main cause that leads to the labels as a ptSTL formula utilizing the temporal semantics. 

The STL formula synthesis problem has been studied in different forms, including finding a formula satisfied by all system traces~\citep{miningjournal,Jha2019}, finding a formula that differentiates the good and the bad traces~\citep{Bartocci2014,Kong:Inference:2014,Bombara:2016}, and, as in our case, finding a formula that identifies the unexpected (or bad) behaviors as soon as they occur~\citep{codit2018}. In particular, in~\citep{codit2018}, as in this work, the exact time of the bad events is integrated to the dataset, whereas, in other works, each trace has a single label. The nuance in the dataset diversify the problem and the solution approaches. 
\cite{miningjournal} presents a method to find valuations for a template STL formula, in which some of the numeric constants are represented with parameters~\citep{Asarin:2011}. Whereas in later works,~\citep{Kong:Inference:2014,Bombara:2016,codit2018} both the structure and the parameters of the formula are generated in an automated way, which eliminates the need for expert guidance. 

In this paper, we propose an iterative framework to synthesize ptSTL formulas in an efficient way. The proposed framework performs parameter synthesis for each parametric ptSTL formula up to a given formula length, and then combines optimized formulas to describe the labels in the given dataset. Thus, it is does not require expert guidance. For parameter synthesis, we present a novel approach that exploits the monotonicity properties of the parameters.
The parameter monotonicity was first introduced by~\cite{miningjournal}, where a tight value is assigned to each parameter in the given order so that the resulting formula is satisfied by each trace. Here, we do not require an ordering between parameters. We generate the parameters that optimize the given monotone criteria in an efficient way, and improve the result through the iterative process.  \cite{Kong:Inference:2014}  presents an iterative formula generation method for a sub-class of STL formulas. As in our method, it iteratively improves the resulting formula. In addition to the differences arising from the variation on the datasets, we consider any STL formula, not only a sub-class, and combine them in a specific way. Furthermore, the proposed iterative synthesis approach allows us to synthesize complex formulas efficiently, which was not possible in~\citep{codit2018}.

The paper is organized as follows. Background information on  STL is given in  Sec.\ref{sec:preliminary}. In Sec.~\ref{sec:problem} , the formula synthesis problem is explained.  The proposed synthesis method is presented in Sec.~\ref{sec:formula_synthesis} and a case study is given in Sec.~\ref{sec:case_studies}. Finally, the closing remarks are given in Sec.~\ref{sec:conclusion}

\section{Preliminaries} \label{sec:preliminary}

\subsection{Signals}
We define an n-dimensional discrete  signal  $\mathbf{x}$ as a mapping from time domain  $\mathbb{N}^+$ to the real numbers $\mathbb{R}^n$.  A finite signal of length $K+1$ is shown as a sequence $\mathbf{x} = x_0, x_1, ...,x_K$. We use $x_t^i$ to denote the projection of the state on the $i$th dimension at time $t$. 

A dataset of labeled signals is defined as 

\begin{align}\label{eq:dataset}
 \mathcal{D} = \{  (\mathbf{x},\mathbf{l}) \mid  \mathbf{x} = & x_0, x_1, \ldots,x_K,  \\
  \mathbf{l} = & l_0, l_1, \ldots, l_K,  \text{ and }  \nonumber \\
  x_t \in &\mathbb{R}^n, l_t \in \{0,1\}, t=0,\ldots,K\}, \nonumber
\end{align}
where $l_t = 1$, means that at time $t$, an event of interest is occurred on signal $\mathbf{x}$.

\subsection{Past Time Signal Temporal Logic}
A Past Time Signal Temporal Logic (ptSTL) formula is defined with grammar:
\begin{align}
\phi = \mathbf{T} | x^i\sim c | \neg\phi | \phi_1 \wedge \phi_2 | \phi_1 \vee \phi_2 | \phi_1 \mathbf{S}_{[a,b]} \phi_2 | \mathbf{P}_{[a,b]} \phi | \mathbf{A}_{[a,b]}  \phi \label{eq:ptstl} 
\end{align}
where $x^i$ is a signal variable, $\sim \in \{>,<\}$,  and $c$ is a constant, $\mathbf{T}$ is the Boolean constant $true$, $\neg, \wedge$ and $\vee$ are the standard Boolean operators, $\mathbf{S}_{[a,b]}$ $(since)$, $\mathbf{P}_{[a,b]}$ $(previously)$, and $\mathbf{A}_{[a,b]}$ $(always)$ are the temporal operators with time interval $[a,b]$. The semantics of a ptSTL formula is defined over a signal for a given time point. 

Informally, for signal $\mathbf{x}$, at time $t$, formula $\mathbf{P}_{[a,b]} \phi $ is satisfied if $\phi$ holds at some time in $[t-b,t-a]$, formula $\mathbf{A}_{[a,b]} \phi$ is satisfied if $\phi$ holds everywhere in $[t-b,t-a]$, and $\phi_1 \mathbf{S}_{[a,b]} \phi_2$ is satisfied if $\phi_2$ holds at some time $t' \in [t-b,t-a]$ and $\phi_1$ holds since then. $(\mathbf{x},t) \models \phi $ denotes that signal $\mathbf{x}$ satisfies formula $\phi$ at time $t$. Formally, the semantics are given as follows:
\begin{align}
& (\mathbf{x}, t) \models  \mathbf{T}  & & \nonumber \\
& (\mathbf{x}, t) \models x^i \sim c  &\text{ iff } &  x^i_t \sim c , \sim \in \{>,<\}\nonumber \\
& (\mathbf{x}, t) \models \phi_1 \wedge \phi_2  & \text{ iff }  & (\mathbf{x}, t) \models \phi_1  \text{ and } (\mathbf{x}, t) \models \phi_2 \nonumber \\
& (\mathbf{x}, t) \models \phi_1 \vee \phi_2  & \text{ iff }  & (\mathbf{x}, t) \models \phi_1  \text{ or } (\mathbf{x}, t) \models \phi_2 \nonumber \\
& (\mathbf{x}, t) \models \mathbf{P}_{[a,b]} \phi & \text{ iff }  & \exists t' \in I(t,[a,b]),  (\mathbf{x}, t') \models \phi \label{eq:semantics} \\
& (\mathbf{x}, t) \models \mathbf{A}_{[a,b]} \phi & \text{ iff }  & \forall t' \in I(t,[a,b]),  (\mathbf{x}, t') \models \phi  \nonumber \\
& (\mathbf{x}, t) \models \phi_1 \mathbf{S}_{[a,b]} \phi_2 & \text{ iff }  & \exists  t' \in I(t,[a,b]),  (\mathbf{x}, t') \models \phi_2, \nonumber \\
& & & \forall t'' \in [t', t]  (\mathbf{x}, t'') \models \phi_1 \nonumber, 
\end{align}
\[ \text{ where } I(t,[a,b]) = [t-b, t-a] \cap [0,t]  \]

Note that the previously ($\mathbf{P}$) and always ($\mathbf{A}$) operators are the special cases of the since operator, $\mathbf{P}_{[a,b]} \phi := \mathbf{T}\ \mathbf{S}_{[a,b]} \phi$ and  $\mathbf{A}_{[a,b]} \phi := \neg \mathbf{P}_{[a,b]} \neg \phi$. We include them as they are used in the proposed methods.

$Parametric$ $Past$ $Time$ $Signal$ $Temporal$ $Logic$ is an extension of ptSTL~\citep{Asarin:2011}. In a  parametric ptSTL formula, instead of numerical values in time interval bounds and predicates, parameters can be used. A parametric formula can be converted to a ptSTL formula by assigning a value to each parameter.
As an example consider the parametric formula $\phi = \mathbf{P}_{[p_1,p_2]} x < p_3$ with parameters $p_1, p_2$ and $p_3$.  ptSTL formula $\phi (v)  = \mathbf{P}_{[3,5]} x < 10.2$ is obtained with valuation $v = [3,5,10.2]$.
 
\subsection{Monotonicity of Parametric Signal Temporal Logic}

Monotonicity properties for parametric STL is introduced by~\cite{miningjournal}. A parametric STL formula $\phi$ with parameters $[p_1, \ldots, p_m]$ is  $monotonically$ $increasing$ with parameter $p_i$ if (\ref{eq:mon_inc}) holds along any signal $\mathbf{x}$. Similarly,  it is $monotonically$ $decreasing$ with parameter $p_i$ if (\ref{eq:mon_dec}) holds.
\begin{align}
&\text{for all }  v,v' \text{ with } v(p_i) < v'(p_i), v(p_j) = v(p_j) \text{ for each } i\neq j, \nonumber \\
& \quad  \quad  \quad (\mathbf{x},t) \models \phi(v) \implies  (\mathbf{x},t) \models \phi(v')   \label{eq:mon_inc} \\
&\text{for all }  v,v' \text{ with }   v(p_i) > v'(p_i),  v(p_j) = v(p_j) \text{ for each } i\neq j,  \nonumber \\
& \quad  \quad  \quad  (\mathbf{x},t) \models \phi(v) \implies  (\mathbf{x},t) \models \phi(v')   \label{eq:mon_dec}
\end{align}

Essentially, $\phi$ is monotonically increasing with $p_i$ if the valuation can not change from satisfying to violating when only the value of the parameter $p_i$ is increased.

Our aim in this work is to generate a ptSTL formula that represents the labels in a dataset~\eqref{eq:dataset}. For this purpose, we generate label $\mathbf{l}^\phi= l^\phi_0, l^\phi_1, \ldots, l^\phi_N$ from a given signal $\mathbf{x} = x_0, \ldots, x_K$ using a given ptSTL formula $\phi$ as follows:
\begin{align}\label{eq:label_set}
	l^\phi_t = &\begin{cases}  1 \text{ if } (\mathbf{x}, t) \models \phi \\
	 0 \text{ otherwise}
	\end{cases}
\end{align}

We define number of positive labels $P^{\#}(\phi, \mathbf{x} )$ (\ref{eq:positive}) and number of negative labels $N^{\#}(\phi, \mathbf{x} )$ (\ref{eq:negative}) where $\mathbf{l}^\phi$ is generated by evaluating  formula $\phi$ along signal $\mathbf{x}$ as defined in~\eqref{eq:label_set}:
\begin{align}\label{eq:positive}
 P^{\#}(\phi, \mathbf{x}) = \sum_{i=0}^{K} l_i^\phi
 \end{align}
\begin{align}\label{eq:negative}
 N^{\#}(\phi, \mathbf{x}) = \sum_{i=0}^{K} \neg l_i^\phi
 \end{align}

 

Also note that
\begin{equation}\label{eq:totalformula}
    P^{\#}(\phi, \mathbf{x}) +  N^{\#}(\phi, \mathbf{x}) = K + 1
\end{equation}

We derive monotonicity properties of $P^{\#}(\phi, \cdot)$ for a parametric ptSTL formula $\phi$ with respect to the monotonicity of $\phi$.  $P^{\#}(\phi,\cdot)$ is monotonically increasing with $p_i$ if and only if the satisfaction value of $\phi$ is monotonically increasing with $p_i$, i.e.,if~\eqref{eq:mon_inc} holds along any signal $\mathbf{x}$, then~\eqref{eq:monoton_inc} holds:
\begin{align}
&\text{for all }  v,v' \text{ with } v(p_i) < v'(p_i), v(p_j) = v(p_j) \text{ for each } i\neq j, \nonumber \\
& \quad  \quad  \quad P^{\#}(\phi(v),\mathbf{x}) \leq P^{\#}(\phi(v'),\mathbf{x})   \label{eq:monoton_inc}
\end{align}
Similarly, if $\phi$ is monotonically decreasing with $p_i$, then  $P^{\#}(\phi, \cdot)$ is also monotonically decreasing with $p_i$.  Specifically, for any signal $\mathbf{x}$,~\eqref{eq:monoton_dec} holds when  \eqref{eq:mon_dec} holds:
\begin{align}
&\text{for all }  v,v' \text{ with } v(p_i) > v'(p_i), v(p_j) = v(p_j) \text{ for each } i\neq j, \nonumber \\
& \quad  \quad  \quad P^{\#}(\phi(v),\mathbf{x}) \leq P^{\#}(\phi(v'),\mathbf{x})   \label{eq:monoton_dec}
\end{align}

Note that, by~\eqref{eq:totalformula}, $P^{\#}(\phi, \cdot)$ and $N^{\#}(\phi, \cdot)$ have the opposite monotonicity property, e.g., if $P^{\#}(\phi, \cdot)$ is monotonically increasing with $p_i$ than $N^{\#}(\phi, \cdot)$ is monotonically decreasing with $p_i$.

In our work, a parameter appears only once in a parametric ptSTL formula. Therefore, the considered formulas are monotonic in each parameter, i.e., either monotonically increasing or monotonically decreasing.

\section{Problem Formulation} \label{sec:problem}


 In this work, our goal is to find a ptSTL formula that represents the labeled events in a dataset~\eqref{eq:dataset}: 

\begin{problem}\label{prob:main} Given a dataset $\mathcal{D}$ as in~\eqref{eq:dataset}, find a ptSTL formula $\phi$ such that for any $(\mathbf{x},\mathbf{l}) \in \mathcal{D}$ and $t \in [0, K]$,  $l_t = l^\phi_t$, where $l^\phi_t$ is as defined in~\eqref{eq:label_set}.
\end{problem}

An exact solution Prob.~\ref{prob:main} is a ptSTL formula $\phi$ that generates the correct label at each time point along each signal in the given dataset (e.g. for each of the $| \mathcal{D} | \times (K+1) $ evaluations). Such an exact formula might not exists due to the noise in the signal and the errors occurred in the labeling process. Here, we aim at finding the \textit{best} formula representing the given dataset. Considering that a number of different reasons can lead to the occurrence of these events, (thus the labels), we iteratively construct a formula via disjunction: 
\begin{equation}\label{eq:endformula}
\phi = \phi_1 \vee \phi_2 \vee \ldots \vee \phi_p.
\end{equation} In particular, in each iteration, we synthesize formula $\phi_i$ with a high number of true positives and bound the number of false positives, since they will propagate via the disjunction operator. In the formula synthesis phase, we consider all parametric formulas with a given number of operators and perform parameter optimization for each formula. For parameter optimization, we present a novel method that exploits the monotonicity of the parameters.

\begin{example}
\label{ex:formulation}
As a running example, we use Aircraft Longitudinal Flight Control model from Simulink~\citep{MATLAB:20106}. In the model, the Pilot's stick pitch command ($pilot$) is the input and it is set as the target point.  Internal commands are generated according to the aircraft's pitch angle ($alpha$) and the pitch rate to reach that target point. Furthermore, the system is perturbed with Dryden wind gust model that feeds gust values $qGust$ and $wGust$ to aircraft dynamics (a sub-system). A trace of the system is shown in Fig~\ref{fig:pilot_wind_alpha}.

\begin{figure}
\begin{center}
{
\resizebox{8.4 cm}{5 cm}{%
\begin{tikzpicture}
\pgfplotsset{
    scale only axis,
}
\begin{axis}[
    axis y line*=left,
    xmin = 0, 
     xmax = 60, 
    xlabel=time (sec),
    ylabel=$pilot$ $alpha$ and $qGust$ ,
]
      
    \addplot[
    color=green, line width = 0.04 cm
    ]
   table {plot_data/aircraft_pilot};
   \label {aircraft_pilot}
    \addlegendentry{data}
         
    \addplot[
    color=blue, line width = 0.04 cm
    ]
   table {plot_data/aircraft_alpha_wind};
   \label {aircraft_alpha_wind}
    \addlegendentry{data}
    
    \addplot[
    color=black, line width = 0.04 cm
    ]
   table {plot_data/windq};
   \label {windq}
    \addlegendentry{data}

\end{axis}  

\begin{axis}[
    axis y line*=right,
    axis x line=none,
    xmin = 0, 
    xmax = 60, 
    ylabel=$wGust$,
]

\addlegendimage{/pgfplots/refstyle=aircraft_pilot}\addlegendentry{$pilot$}
\addlegendimage{/pgfplots/refstyle=aircraft_alpha_wind}\addlegendentry{$alpha$}
\addlegendimage{/pgfplots/refstyle=windq}\addlegendentry{$qGust$}
 \addplot[
    color=red, line width = 0.04 cm
    ]
   table {plot_data/windw};
   \label {windw}

\addlegendentry{$wGust$}
\end{axis}  
\end{tikzpicture}
}

}
\caption{Aircraft Longitudinal Flight Control Example} 
\label{fig:pilot_wind_alpha}
\end{center}
\end{figure}
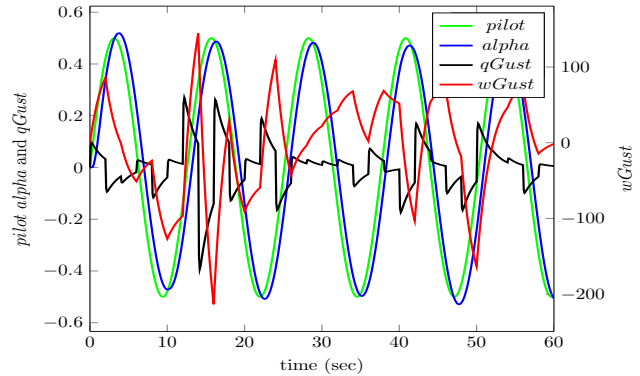

 In this example, our aim is to find the conditions which cause the aircraft's longitudinal motion to disturb. For this purpose, we run two different scenarios. In the first scenario, we set noise power of the band-limited white noise generator ($noise$) in the Dryden wind gust model to 0 and collect the $alpha$ output of the model, denoted as $alpha^0$. In the second scenario, we set noise power of  band-limited white noise generator to 1000, set sample time of the noise generator to 2 secs and again collect the $alpha$ output of the model, denoted as $alpha^1$. In both experiments, the same input signal ($pilot$) is used. The difference is computed along $alpha^0$ and $alpha^1$. This difference can be seen as how much the aircraft is disturbed from the normal behavior. The following formula is used to label the traces with high noise (second case).
\begin{align}\label{eq:label_formula}
 |alpha^0 - alpha^1| > 0.06
\end{align}
Time points at which this formula is satisfied are labeled with 1, and others are labeled with 0. We generate dataset  $\mathcal{D}$ with 5 labeled traces each having a length of 600. Note that seed parameter of the Band-Limited White Noise is set to $k$ for $k^{th}$ trace. Each data point includes the pilot's command, aircraft's pitch angle and the wGust and qGust outputs of the Dryden Wind Gust Model in which we set the noise power to 1000:  
\begin{align}\label{lb:dataset_example}
    x_i = \{pilot_i , alpha^1_i, wGust_i, qGust_i\} 
\end{align}
Out of 3000 data points, 477 of them are labeled with 1 and the rest is labeled with 0. An example of the difference  $( alpha^0 - alpha^1 )$ and the label are shown in Fig. \ref{fig:labeled_trace}.

\begin{figure}
\begin{center}
{
\resizebox{8.4 cm}{5 cm}{%
\begin{tikzpicture}
\pgfplotsset{
    scale only axis,
}
\begin{axis}[
    axis y line*=left,
    xmin = 0, 
    xmax = 60, 
    xlabel=time (sec),
    ylabel=$alpha^0 - alpha^1$,
]
      
    \addplot[
    color=green,line width = 0.04 cm
    ]
   table {plot_data/aircraft_diff};
   \label {aircraft_diff}
    \addlegendentry{data}

\end{axis}  

\begin{axis}[
    xmin = 0, 
    xmax = 60, 
    axis y line*=right,
    axis x line=none,
    ylabel= $label$,
]

\addlegendimage{/pgfplots/refstyle=aircraft_diff}\addlegendentry{$alpha^0 - alpha^1$}
 \addplot[
    color=red,line width = 0.04 cm
    ]
   table {plot_data/aircraft_label};
   \label {aircraft_label}

\addlegendentry{$label$}
\end{axis}  
\end{tikzpicture}
}
}
\caption{$alpha^0 - alpha^1$ and generated label $l$} 
\label{fig:labeled_trace}
\end{center}
\end{figure}
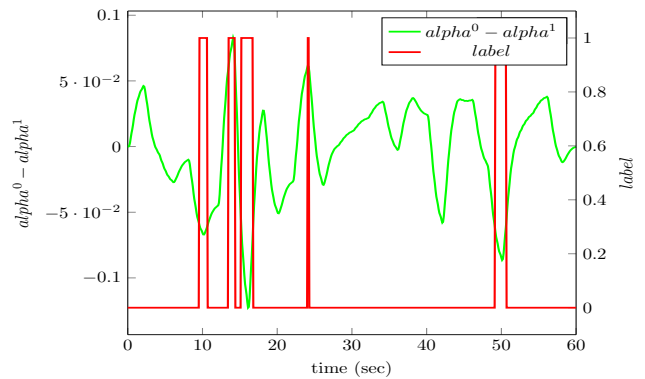

\end{example}

\section{Formula Synthesis Method}\label{sec:formula_synthesis}

In this section, we present the proposed formula synthesis method to solve Prob.~\ref{prob:main}. First, we define success measures for the ptSTL formulas over the considered dataset $\mathcal{D}$, and derive monotonicity properties for these with respect to formula parameters. Then, we propose an efficient method based on monotonicity properties to find parameters of a given parametric ptSTL formula. Finally, we present the developed iterative formula synthesis approach which utilizes the parameter synthesis method for efficient computation. 

\subsection{Monotonicity for Temporal Parameters}
The number of positive labels $P^{\#}(\phi, \mathcal{D})$ of a ptSTL formula $\phi$ over a dataset $\mathcal{D}$ is simply defined as the total number of positive labels, and derived from~\eqref{eq:positive}. $N^{\#}(\phi, \mathcal{D})$ is defined similarly.
\[ P^{\#}(\phi, \mathcal{D}) = \sum_{(\mathbf{x}, \mathbf{l}) \in \mathcal{D}}^{K} P^{\#}(\phi, \mathbf{x})  \quad N^{\#}(\phi, \mathcal{D}) = \sum_{(\mathbf{x}, \mathbf{l}) \in \mathcal{D}}^{K} N^{\#}(\phi, \mathbf{x})\]

As either a positive ($1$) or a negative ($0$) label is assigned to each data point, the equality $|\mathcal{D}| \times (K+1) = P^{\#}(\phi, \mathcal{D})  + N^{\#}(\phi, \mathcal{D})$ trivially holds. We define the number of correctly identified positive instances (\textit{true positives}) with respect to the labels generated by the formula $\phi$ using~\eqref{eq:label_set} and the dataset labels as:
\begin{align}\label{eq:tp}
  TP^{\#}(\phi, \mathcal{D}) = \sum_{(\mathbf{x},\mathbf{l}) \in \mathcal{D}}  \sum_{i=0}^{K} l_i \wedge l^\phi _i
 \end{align}
 
Similarly, the total number of incorrect positive results, i.e., the data points that have label $0$ in the given dataset and label $1$ according to the ptSTL formula $\phi$ ($l_i^\phi=1$) is defined as:
\begin{align}\label{eq:fp}
 FP^{\#}(\phi, \mathcal{D}) = \sum_{(\mathbf{x},\mathbf{l})\in \mathcal{D}}  \sum_{i=1}^{K}  \neg l _i \wedge l^\phi_i 
 \end{align}
 
The derivations of $TP^{\#}(\cdot, \cdot)$ and $FP^{\#}(\cdot, \cdot)$ preserves monotonicity properties~\eqref{eq:monoton_inc} and~\eqref{eq:monoton_dec}. Therefore, if a parametric ptSTL formula $\phi$ is  increasing (or decreasing) with a parameter $p$, then both $TP^{\#}(\cdot, \cdot)$ and $FP^{\#}(\cdot, \cdot)$ are increasing (or decreasing) with $p$. 

We use $\mathcal{M}(p,\phi)$ to denote the monotonicity property of parameter $p$ in $\phi$ for the number of positives ($TP^{\#}(\cdot, \cdot)$, $FP^{\#}(\cdot, \cdot)$ or $P^{\#}(\cdot,\cdot)$ ):
\begin{align}\label{eq:monotonicity}
	\mathcal{M}(p,\phi) = &\begin{cases}
	 \mathbf{I}  \text{ if } p \text{ is monotonically increasing in } \phi  \\
	 \mathbf{D} \text{ if } p \text{ is monotonically decreasing in } \phi
	\end{cases}
\end{align}

Monotonicity property, $\mathcal{M}(\cdot,\cdot) $, for each parameter in a basic formula is given in Table~\ref{tb:monotonicity_table}. 
 
\begin{table}[h]
\begin{center}
\caption{\textsc{Monotonicity Table}}\label{tb:monotonicity_table}
\begin{tabular}{cc|cc}
$\phi$   & $\mathcal{M}(p,\phi)$ &   $\phi$  & $\mathcal{M}(p,\phi)$ \\ \hline\hline
$x > p$  &   $\mathbf{D}$ & $x < p$  &   $\mathbf{I}$\\ \hline
$\mathbf{A}_{[c, p]} \varphi$  & $\mathbf{D}$ & $\mathbf{A}_{[p, c]} \varphi$  & $\mathbf{I}$\\ \hline
$\mathbf{P}_{[c, p]}\varphi$  & $\mathbf{I}$ & $\mathbf{P}_{[ p, c]}\varphi$  & $\mathbf{D}$  \\ \hline
$\varphi_1 \mathbf{S}_{[c, p]}\varphi_2$  & $\mathbf{I}$ & $\varphi_1 \mathbf{S}_{[p,c]}\varphi_2$   & $\mathbf{D}$  \\ \hline
\end{tabular}
\end {center}
\end{table}

Note that the preceding derivations are based on the number of positive labels. The number of correctly identified negative labels, $TN^{\#}(\phi, \mathcal{D})$, and the number of incorrectly identified negative labels $FN^{\#}(\phi, \mathcal{D})$ are defined similarly. These show the opposite monotonicity property, i.e., if $TP^{\#}(\phi, \mathcal{D})$ is monotonically increasing in parameter $p$, then $TN^{\#}(\phi, \mathcal{D})$ is monotonically decreasing in $p$. 
Furthermore, the negation operator ($\neg$) inverts the monotonicity property. For example, while $\mathcal{M}(p,\mathbf{P}_{[a,b]} x < p)$ is $\mathbf{I}$, $\mathcal{M}(p,\neg \mathbf{P}_{[a,b]} x < p)$ is $\mathbf{D}$. A parameter's monotonicity is determined by checking the syntax tree of the formula: each negation that appears from the root node to the parameter inverts the monotonicity of the parameter shown in Table~\ref{tb:monotonicity_table}.

\begin{example}
\label{ex:pptstl_formula}

Consider parametric ptSTL formula
\begin{equation}\label{eq:exformula}
\phi =  ( \mathbf{P}_{[p_1, p_2]} ( qGust < p_3 ) )  \wedge  ( wGust < p_4 )
 \end{equation} 
Monotonicity properties of $p_1, p_2, p_3$ and  $p_4$ are $
\mathcal{M}(p_1,\phi) = \mathbf{D}$, $
\mathcal{M}(p_2,\phi) = \mathbf{I}$, $
\mathcal{M}(p_3,\phi) = \mathbf{I}$, $
\mathcal{M}(p_4,\phi) = \mathbf{I}.$
\end{example}

\subsection{Parameter Optimization Using Monotonicity}\label{sec:parametersynthesis}

We now present an efficient method based on monotonicity to find parameters of a parametric ptSTL formula $\phi$ from a given dataset $\mathcal{D}$~\eqref{eq:dataset} such that the number of correctly identified positives of the resulting formula is maximized while the number false positives is below a given threshold:  

\begin{problem}\label{prob:optimization} 
Given a labeled dataset $\mathcal{D}$~\eqref{eq:dataset}, a parametric ptSTL formula $\phi$ with $n$ parameters $p_1, p_2, \ldots, p_n$, lower and upper bounds $l_i, u_i$ for each parameter $p_i$, an error bound $B \in \mathbb{N}$, find the valuation $v$  within the given limits that maximizes $TP^{\#}(\phi(v), \mathcal{D})$ while guaranteeing that $ FP^{\#}(\phi(v), \mathcal{D}) \leq B$. 

\end{problem}

To solve this problem, we first present an algorithm for parametric ptSTL formulas with two parameters, and then discuss how this approach is adapted for parametric ptSTL formulas with more than two parameters.

We present a diagonal search method to solve Prob.~\ref{prob:optimization} in an efficient way when $n$ is 2, which adapts search problem of the product of an m element chain and an n element chain~\citep{linial1985searching} for ptSTL parameter optimization. The diagonal search algorithm starts with a valuation $v$ with $v(p_1)$ is the bound on $p_1$ that maximizes $TP^{\#}(\phi, \mathcal{D})$ (i.e. either $l_1$ or $u_1$) and $v(p_2)$ is the bound on $p_2$ that minimizes $TP^{\#}(\phi, \mathcal{D})$. Given step sizes $\delta_1$ and $\delta_2$ for both parameters, the algorithm iteratively changes the value of a parameter according to the following rule: change $v(p_1)$ by $\delta_1$ in the direction decreasing $P^{\#}(\phi, \mathcal{D})$ if the error constraint does not hold at $v$, otherwise change $v(p_2)$ by $\delta_2$ in the direction increasing $P^{\#}(\phi, \mathcal{D})$. Thus, the algorithm moves along a diagonal of the product of the discretized parameter domains with the objective of satisfying the error bound or improving the optimization criteria. This diagonal method is summarized in Alg.~\ref{alg:diagonal}

\begin{algorithm}
\caption{$DiagonalSearch(\phi,B,\mathcal{D}, l_1, u_1, \delta_1, l_2, u_2, \delta_2)$}\label{alg:diagonal}
\begin{flushleft}
\begin{algorithmic}[1]
\Require{$\phi$: A parametric ptSTL formula with parameters $p_1$ and $p_2$, $B$ : bound on $FP^{\#}(\phi(v), \mathcal{D})$, $\mathcal{D}$: dataset as in~\eqref{eq:dataset},  $l_i, u_i, \delta_i$: lower bound, upper bound and step size for parameter $p_i$, $i \in \{1,2\}$}
\Ensure{$v_{best} = \arg\max_{  v } \{TP^{\#}(\phi(v),\mathcal{D}) \mid FP^{\#}(\phi(v),\mathcal{D}) < B\}$} 
\If {$\mathcal{M}(p_1,\phi) == \mathbf{I}$} \label{line:startinitialize}
	\State $v(p_1) = u_1, \bar \delta_1 = -\delta_1$
\Else
\State  $v(p_1) = l_1, \bar \delta_1 = \delta_1$
\EndIf
\If {$\mathcal{M}(p_2,\phi) == \mathbf{I}$}
	\State $v(p_2) = l_1, \bar \delta_2 = \delta_2$
\Else
\State  $v(p_2) = u_1, \bar \delta_2 = -\delta_2$
\EndIf  \label{line:endinitialize}
\State $v_{best} = []$,  $TP_{best} = 0$
\While{$l_1 \leq v(p_1) \leq u_1 \wedge l_2 \leq v(p_2) \leq u_2$}\label{line:loopstart}
    \If{$B < FP^{\#}(\phi(v),\mathcal{D})$} \label{line:errorconstraint}
        \State $v(p_1) = v(p_1) + \bar \delta_1$
    \Else
        \If{$TP^{\#}(\phi(v),\mathcal{D}) \geq TP_{best}$}\label{line:bestknown}
                \State $TP_{best} = TP^{\#}(\phi(v),\mathcal{D})$, $v_{best} = v$
        \EndIf
        \State $v(p_2) = v(p_2) + \bar \delta_2$
    \EndIf
\EndWhile\label{line:loopend}
\State \Return $v_{best}$
\end{algorithmic}
\end{flushleft}
\end{algorithm}

In lines~\ref{line:startinitialize}-\ref{line:endinitialize} of Alg.~\ref{alg:diagonal}, the initial value and the update direction is defined for each parameter with respect to its monotonicity property. At each iteration of the main loop (lines~\ref{line:loopstart}-\ref{line:loopend}), exactly one parameter value is updated. If the error constraint (line~\ref{line:errorconstraint}) is violated, the parameter that initialized to maximize $TP^{\#}(\phi,\mathcal{D})$, $p_1$, is changed by $\bar \delta_1$ to reduce $FP^{\#}(\phi(v),\mathcal{D})$. Otherwise, the current parameter assignment is a candidate solution, and it is checked against the best known solution (line~\ref{line:bestknown}). Then, the parameter that initialized to minimize $FP^{\#}(\phi,\mathcal{D})$, $p_2$, is changed by $\bar \delta_2$ to increase $TP^{\#}(\phi(v),\mathcal{D})$. The iterations end when a parameter is out of the given bounds. Consequently, $O(m_1 + m_2)$ formula evaluations are performed over the given dataset, where $m_1 = \frac{u_1 - l_1}{\delta_1},  m_2 = \frac{u_2 - l_2}{\delta_2}$.

\begin{figure}
\begin{center}
\resizebox{8.4 cm}{4.2 cm}{%
\begin{tikzpicture}[scale=.6]
\begin{scope}

\fill[black!30!red,opacity=0.6]    (3,0) rectangle (5,1);
\fill[black!30!red,opacity=0.6]    (5,0) rectangle (7,5);
\fill[black!30!red,opacity=0.6]    (6,5) rectangle (7,6);

\fill[black!30!green,opacity=0.6]    (1,0) rectangle (3,6);
\fill[black!30!green,opacity=0.6]    (3,1) rectangle (5,6);
\fill[black!30!green,opacity=0.6]    (5,5) rectangle (6,6);
\draw (1, 0) grid (7, 6);

\draw[very thick, scale=1] (1, 0) grid (3, 0);
\draw[very thick, scale=1] (3, 0) grid (3, 1);
\draw[very thick, scale=1] (3, 1) grid (5, 1);
\draw[very thick, scale=1] (5, 1) grid (5, 5);
\draw[very thick, scale=1] (5, 5) grid (6, 5);
\draw[very thick, scale=1] (6, 5) grid (6, 6);
\draw[very thick, scale=1] (6, 6) grid (7, 6);
\tikzset{anchor=west}

\setcounter{row7}{1}
\setrowseven {}{\begin{turn}{90}\tiny $\: \: \:\: \:l_1$\end{turn}}{\begin{turn}{90} \tiny $\: \: \:\: \:   l_1 + \delta_1$\end{turn}}{\tiny ..}{\tiny .}{\begin{turn}{90}\tiny $ \: \: \:\: \:u_1 - \delta_1$\end{turn}}{\begin{turn}{90}\tiny $\: \: \:\: \:u_1$\end{turn}}
\setrowseven {\begin{turn}{-45}\tiny $l_2 $\end{turn}}{1}{2}{2}{2}{3}{5}
\setrowseven {\begin{turn}{-45}\tiny $l_2 + 1\delta_2$\end{turn}}{1}{2}{2}{3}{4}{5}
\setrowseven {\begin{turn}{-45}\tiny $l_2 + 2\delta_2$\end{turn}}{1}{2}{3}{3}{4}{5}
\setrowseven {\begin{turn}{-45}\tiny $l_2 + 3\delta_2$\end{turn}}{2}{2}{3}{3}{4}{5}
\setrowseven {\begin{turn}{-45}\tiny $l_2 + 4\delta_2$\end{turn}}{2}{2}{3}{3}{4}{5}
\setrowseven {\begin{turn}{-45}\tiny $l_2 + 5\delta_2$\end{turn}}{3}{3}{4}{4}{4}{5}

\node[anchor=center] at (4.0, -0.5) {$FP^{\#}(\phi(v), \mathcal{D})$};
\end{scope}

\begin{scope}[xshift=7.5cm]

\fill[black!30!red,opacity=0.6]    (3,0) rectangle (5,1);
\fill[black!30!red,opacity=0.6]    (5,0) rectangle (7,5);
\fill[black!30!red,opacity=0.6]    (6,5) rectangle (7,6);

\fill[black!30!green,opacity=0.6]    (1,0) rectangle (3,6);
\fill[black!30!green,opacity=0.6]    (3,1) rectangle (5,6);
\fill[black!30!green,opacity=0.6]    (5,5) rectangle (6,6);
\draw (1, 0) grid (7, 6);

\draw[very thick, scale=1] (1, 0) grid (3, 0);
\draw[very thick, scale=1] (3, 0) grid (3, 1);
\draw[very thick, scale=1] (3, 1) grid (5, 1);
\draw[very thick, scale=1] (5, 1) grid (5, 5);
\draw[very thick, scale=1] (5, 5) grid (6, 5);
\draw[very thick, scale=1] (6, 5) grid (6, 6);
\draw[very thick, scale=1] (6, 6) grid (7, 6);

\draw[line width=2pt,<-] (5.5,5.5)--(6.5,5.5) node[right]{};
\draw[line width=2pt,<-] (5.5,4.5)--(5.5,5.5) node[right]{};
\draw[line width=2pt,<-] (4.5,4.5)--(5.5,4.5) node[right]{};
\draw[line width=2pt,<-] (4.5,0.5)--(4.5,4.5) node[right]{};
\draw[line width=2pt,<-] (2.5,0.5)--(4.5,0.5) node[right]{};
\draw[line width=2pt,<-] (2.5,0)--(2.5,0.5) node[right]{};
\tikzset{anchor=west}

\setcounter{row7}{1}

\setrowseven {}{\begin{turn}{90}\tiny $\: \: \:\: \:l_1$\end{turn}}{\begin{turn}{90} \tiny $\: \: \:\: \:   l_1 + \delta_1$\end{turn}}{\tiny ..}{\tiny .}{\begin{turn}{90}\tiny $ \: \: \:\: \:u_1 - \delta_1$\end{turn}}{\begin{turn}{90}\tiny $\: \: \:\: \:u_1$\end{turn}}
\setrowseven {\begin{turn}{-45}\tiny $l_2$\end{turn}}{10}{11}{14} {15}{17} {18}
\setrowseven {\begin{turn}{-45}\tiny $l_2 + \delta_2$\end{turn}}{10}{12}{15} {16}{17} {19}
\setrowseven {\begin{turn}{-45}\tiny $l_2 + 2\delta_2$\end{turn}}{11}{12}{15} {17}{18} {21}
\setrowseven {\begin{turn}{-45}\tiny $l_2 + 3\delta_2$\end{turn}}{13}{13}{17} {19}{21} {22}
\setrowseven {\begin{turn}{-45}\tiny $l_2 + 4\delta_2$\end{turn}}{14}{14}{18} {20}{22} {24}
\setrowseven {\begin{turn}{-45}\tiny $l_2 + 5\delta_2$\end{turn}}{15}{17}{22} {24}{28} {30}

\node[anchor=center] at (4.0, -0.5) {$TP^{\#}(\phi(v), \mathcal{D})$};
\end{scope}

\end{tikzpicture}
}
\end{center}
\caption{An example run of Alg.~\ref{alg:diagonal}.} \label{fig:grid}
\end{figure}

An example run of Alg.~\ref{alg:diagonal} is shown in Fig.~\ref{fig:grid} for illustration, where $B$ is $3$, both $\mathcal{M}(p_1,\phi)$ and $\mathcal{M}(p_2,\phi)$ are  $\mathbf{I}$. In Fig. \ref{fig:grid}, each cell contains  $FP^{\#}(\phi(v), \mathcal{D})$ and $TP^{\#}(\phi(v), \mathcal{D})$ on the left and right arrays, respectively. Cells with $ FP^{\#}(\phi(v), {\mathcal{D}}) > B$ are marked with red (infeasible parameters) and the rest of the cells are marked with green. The algorithm takes a step to the left when it encounters a red cell and takes a step to the down when it encounters a green cell. The path the algorithm follows is shown on the grid. The algorithm returns $[u_1 - 2\delta_1, l_2 + 4\delta_2]$. Note that the algorithm evaluates the optimal in each row it finds the overall optimal for the given step sizes.

We now describe the proposed method to solve Prob.~\ref{prob:optimization}. Let $\phi$ be the given parametric ptSTL formula with n parameters $p_1, \ldots, p_n$. If $n=1$, the optimal value is found with a binary search. If $n=2$, the optimal value is found with $DiagonalSearch$ method described in Alg.~\ref{alg:diagonal}. Finally, if $n > 2$, $DiagonalSearch$ is run for $p_1$ and $p_2$ for all possible combinations of the last $n-2$ parameters, and the optimal parameters are returned. The whole process is referred as $ParameterSynthesis(\phi, B, \mathcal{D})$.

\begin{example}
\label{ex:optimization}
Consider the parametric ptSTL formula~\eqref{eq:exformula} from Ex. \ref{ex:pptstl_formula} with parameter ranges: $l_1, l_2 = 0$, $u_1,u_2 = 30$, $l_3=-0.4$, $u_3= 0.3$, $l_4=-240$, $u_4=210$.   
$2,0.05,0.05$ and $30$ are set as the step sizes $\delta_1\delta_2,\delta_3$ and $\delta_4$, respectively. Since $n>2$, two of the parameters, namely $p_2, p_3$, are selected for $DiagonalSearch$ based on the size of the parameter domains. 
For the remaining parameters $p_1$ and $p_4$, $\phi^{i,j}$ is created as follows:
\begin{align}
\phi^{i,j} =& \phi(v) \text{ where } v = [i, p_2, p_3 , j ],  \\ \nonumber
 & i \in \{0,2, \ldots 30\}, j \in \{-240, -210, \ldots 210\}
\end{align}
Alg. \ref{alg:diagonal} is run for each $\phi^{i,j}$ with $B=5$ over the dataset $\mathcal{D}$  defined in  Ex.~\ref{ex:formulation}. The maximum $TP^{\#}(\phi^{i,j}(v),\mathcal{D})$ is attained when $i=4$ and $j=-20$ with $v_{best} = [10, 0]$, which corresponds to ptSTL formula 
$\phi(v) = (\mathbf{P}_{[4, 10]}  qGust < 0 )  \wedge  ( wGust < -120 ))$ with $TP^{\#}(\phi,\mathcal{D}) = 235$ and $FP^{\#}(\phi(v),\mathcal{D}) = 4$. This formula explains approximately $50\%$ of the large deviations from the normal behavior (label 1). Thus a possible approach would be adjusting internal commands generated from $pilot$ with respect to the wind behavior characterized by $\phi(v)$.
Note that while grid search requires 61440 valuations~\citep{codit2018}, $ParameterSynthesis(\phi, B, \mathcal{D})$ can find the optimal solution with $4721$ valuations.

\end{example}

\subsection{Formula Synthesis}
In this section, we present the solution to the main problem (Prob.~\ref{prob:main}) considered in this paper: find a ptSTL formula that represents labeled events in  a dataset.
In general, an unexpected behavior/fault can occur due to a number of different reasons. To utilize this property, we iteratively construct a formula for the given dataset $\mathcal{D}$ as a disjunction of ptSTL formulas each representing a different reason. The goal in each iteration is to find a formula for a subset of the labeled instances, while limiting incorrectly labeled instances (FP) as this type of error propagates with disjunction operator. To find such a formula, we define the set of all parametric formulas as in~\citep{codit2018}, and perform parameter optimization on each of them using $ParameterSynthesis$ method described in Sec.~\ref{sec:parametersynthesis}. 

Given the set of system variables, $\{x^1, \ldots,x^n\}$, and a bound on the number of operators $N$, the set of all parametric ptSTL formulas with up to $N$ operators $\mathcal{F}^{\leq N}$ is recursively defined as:
\begin{align}\label{eq:formula_space}
\mathcal{F}^{0} &= \{  x^i \sim p_i \mid  \sim \in \{<,>\}, i = 1,\ldots,n\} \cup \{\mathbf{T}\} \\
\mathcal{F}^{N} &= \{\neg  \phi , \mathbf{P}_{[a,b]} \phi, \mathbf{A}_{[a,b]} \phi \mid \phi \in \mathcal{F}^{N-1} \} \cup \nonumber \\ 
&\bigcup\limits_{i=1}^{n-1} \{ \phi_{1} \wedge \phi_{2} ,  \phi_{1} \vee \phi_{2} , \phi_{1}\mathbf{S}_{[a,b]}\phi_{2}  \mid \nonumber \\
& \quad\quad\quad\quad \phi_{1} \in \mathcal{F}^{i}, \phi_{2} \in \mathcal{F}^{N-i-1}\}  \nonumber \\
\mathcal{F}^{\leq N} &= \cup_{i=0}^{N} \mathcal{F}^{i}\nonumber
\end{align}

\begin{algorithm}
\caption{$FormulaSynthesis(\mathcal{F}, B, \mathcal{D}, p)$}\label{alg:synthesize}  

\begin{algorithmic}[1]
\Require{$\mathcal{F}$: a set of parametric ptSTL formulas, $B$: bound on the number of false positives, $\mathcal{D}$: a dataset as in~\eqref{eq:dataset}, $p$: upper bound on the number of formulas concatenated with disjunction.} 

 \State $\mathcal{F}^v= \{ \phi(v) = ParameterSynthesis( \phi, B, \mathcal{D}) \mid \phi \in \mathcal{F}\}$ \label{line:parametersyn}
 \State $i=0, TP_{prev} = 0, TP=1$, $\Phi = false$
 \While{$TP > TP_{prev}$\text{ and } $i < p$}\label{line:loopstart2}
 \State $\phi(v)^* = \arg \max_{\phi(v) \in \mathcal{F}^v} TP^{\#}(\Phi \vee \phi(v), \mathcal{D})$
 \State $\Phi = \Phi \vee \phi(v)^*$
 \State $i=i+1$

  \State $TP_{prev} = TP$, $TP = TP^{\#}(\Phi, \mathcal{D})$
  \EndWhile\label{line:loopend2}

 \State \Return $\Phi$
\end{algorithmic}
\end{algorithm}

The proposed formula synthesis approach is summarized in Alg.~\ref{alg:synthesize}. The method takes a set of parametric formulas $\mathcal{F}$, a bound on the number of false positives $B$, a labeled dataset $\mathcal{D}$ and a bound $p$ on the number of ptSTL formulas, and generates a ptSTL formula $\phi^\star$ in the form of~\eqref{eq:endformula} with at most $p$ sub-formulas, such that $FP^{\#}(\phi^\star, \mathcal{D}) < B p$ and $TP^{\#}(\phi^\star, \mathcal{D})$ is optimized. The set of parametric formulas can be defined as in~\eqref{eq:formula_space}, or alternatively, an expert of the considered system can write a set of parametric formulas. In the algorithm, first, parameters are optimized for each parametric ptSTL formula $\phi \in \mathcal{F}$ (line~\ref{line:parametersyn}). Then, starting from $\Phi = false$, iteratively, the formula $\phi(v)^\star$ maximizing the valuation of the combined formula $\Phi \vee \phi(v)^\star$ is selected from the set of ptSTL formulas $\mathcal{F}^v$ until the sub-formula limit $p$ is reached, or concatenating new formulas does not improve the result (lines~\ref{line:loopstart2}-\ref{line:loopend2}). Note that at each iteration a formula $\phi(v)^\star$ is added to $\Phi$ with disjunction.

In Alg.~\ref{alg:synthesize}, $ParameterSynthesis( \phi, B, \mathcal{D}) $ is run only once for each parametric ptSTL formula. At every iteration of the algorithm, $TP^{\#}(\Phi \vee \phi(v), \mathcal{D})$ is computed for each $\phi(v) \in \mathcal{F}^v$ to select the formula $\phi(v)^\star$ that generate the highest increment in $TP$. Note that the resulting formula $\phi^\star$ might not be the optimal formula due to the iterative synthesis approach. Essentially, the fitness of the formula is upper bounded by the formula that would be obtained by performing parameter optimization on parametric formulas in the form of~\eqref{eq:endformula} with $N \times p$ parameters (as in~\citep{codit2018}). 
However, due to the complexity of the parameter synthesis algorithm, this computation is not feasible for large formulas.

\begin{example}
\label{ex:synthesis}
The set of all parametric ptSTL formulas $\mathcal{F}^{\leq 2}$ with  at most $2$ parameters over the system variables $\{alpha, pilot, wGust, qGust\}$ is generated according to~\eqref{eq:formula_space}. The parameter domains are defined as:
$p_a, p_b \in \{2i \mid i=0, \ldots,15   \}$  for $\mathbf{A}_{[p_a, p_b]}$,$\mathbf{P}_{[p_a, p_b]}$,  \\
$p_{alpha} \in \{-0.5 + 0.05i \mid i=0, \ldots,20   \}$, \\
$p_{pilot} \in \{-0.5 +05i \mid i=0, \ldots,20   \}$, \\
$p_{wGust} \in \{-240 +30i \mid i=0, \ldots,15   \}$, \\ 
$p_{qGust} \in \{-0.4 - 0.05i \mid i=0, \ldots,14   \}$.  
 
We run Alg.~\ref{alg:synthesize} with the parametric formula set $\mathcal{F}^{\leq 2}$, the dataset from  Ex.~\ref{ex:formulation}, bound $B=5$ and subformula limit $p=4$.  The resulting formula is:

\begin{align}
& \phi = \phi_1 \vee \phi_2 \vee \phi_3  \vee \phi_4 \\  \nonumber
& \phi_1 = ( \mathbf{P} _{[4, 10]} ( qGust < 0 ) ) \wedge ( wGust < -120 )   \\   \nonumber 
& \phi_2 = ( wGust > 120 ) \wedge ( \mathbf{A} _{[14, 14]} ( pilot > -40 ) )  \\  \nonumber 
&  \phi_3 =\mathbf{P} _{[2, 2]} ( ( alpha < 30 ) \wedge ( wGust < -120 ) )\\  \nonumber
& \phi_4 =( \mathbf{A} _{[4, 16]} ( qGust > 10 ) ) \wedge ( pilot < -40 )  \nonumber 
  \end{align}
  Each sub-formula  $\phi_1, \phi_2 $, $\phi_3$ and $ \phi_4$ explains a condition that led to a disturbance in the pitch angle of the aircraft. 
  The first formula $\phi_1$ shows that a disturbance occurs when $wGust$ is less than $-120$ and $qGust$ was lower than $0$ for some time within the last 10 time steps to last 4 time steps. Formulas $\phi_2 $, $\phi_3$ and $ \phi_4$ show that a disturbance occurs 2) when $wGust$ is greater than $120$ and $pilot$ was greater then $-40$ $14$ time steps ago, or 3) if $alpha$ was less than $30$ and $wGust$ was less than -120   two steps ago, or 4) if $qGust$ was higher than 10 for each step between last 4 and 16  steps and $pilot$ is less than -40 in the current step.

 $477$ out of $3000$ data points in $\mathcal{D}$ are labeled with $1$. 
 $TP^{\#}(\phi,\mathcal{D}) $ and $FP^{\#}(\phi,\mathcal{D})$ valuations are 419 and 18 respectively. Total mismatch count of 3000 points is computed as 76 which leads to an accuracy of 97.46\%.

This result is found in 3350 seconds on a PowerEdge T430 machine with Intel Xeon E5-2650 12C/24T processor. It is important to note that $\phi$ includes $11$ operators and $15$ parameters and it is defined over $4$ system variables. This example shows that the proposed method can generate complex formulas from labeled datasets in an efficient way, since, due to the computational complexity, existing formula synthesis algorithms are validated on simpler formulas.

\end{example}

\section{Case Study}\label{sec:case_studies}

\begin{figure}[h]
\begin{center}
\includegraphics[width=8.4cm]{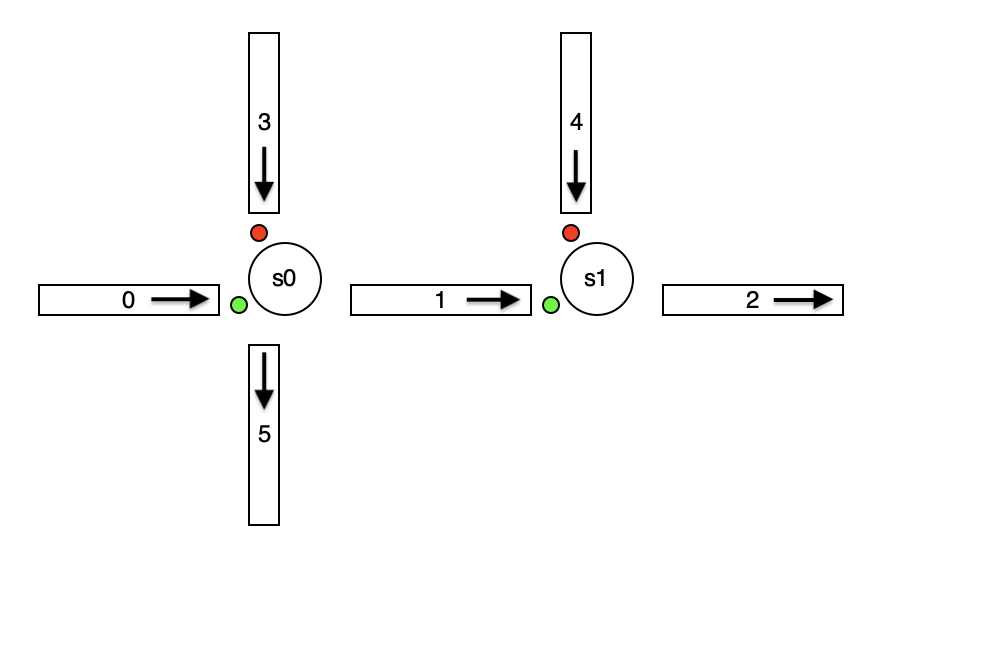}    
\caption{Traffic network containing 2 signals and 6 links.} 
\label{fig:traffic}
\end{center}
\end{figure}

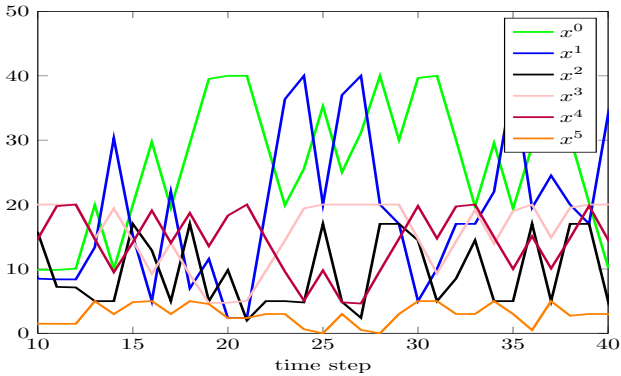
\begin{figure}[h]
\begin{center}
{
\resizebox{8.4 cm}{5 cm}{%
\begin{tikzpicture}
\pgfplotsset{
    scale only axis,
}
\begin{axis}[
    xmin = 10, 
     xmax = 40, 
    ymin = 0, 
     ymax = 50, 
    xlabel=time step,
]   
    \addplot[
    color=green, line width = 0.04 cm
    ]
   table {plot_data/file_traf_x0};
   \label {file_traf_x0}
         
    \addplot[
    color=blue, line width = 0.04 cm
    ]
   table {plot_data/file_traf_x1};
   \label {file_traf_x1}
    
    \addplot[
    color=black, line width = 0.04 cm
    ]
   table {plot_data/file_traf_x2};
   \label {file_traf_x2}
    \addplot[
    color=pink, line width = 0.04 cm
    ]
   table {plot_data/file_traf_x3};
   \label {file_traf_x3}
    \addplot[
    color=purple, line width = 0.04 cm
    ]
   table {plot_data/file_traf_x4};
   \label {file_traf_x4}
    \addplot[
    color=orange, line width = 0.04 cm
    ]
   table {plot_data/file_traf_x5};
   \label {file_traf_x5}
   \legend{$x^0$,$x^1$,$x^2$,$x^3$,$x^4$,$x^5$}

\end{axis}  

\end{tikzpicture}
}

}
\caption{A sample trace of the traffic system shown in Fig.~\ref{fig:traffic}.} 
\label{fig:traf_tra}
\end{center}
\end{figure}

As a case study, we consider a traffic system that consist of  6 links and 2 traffic signals shown in Fig.~\ref{fig:traffic}. The traffic network is modeled as a piecewise affine system. The state vector of the model captures the number of vehicles $x^i$ on each link $i$, and the configuration $s^j$ of each signal $j$ is the system input. The details of the model can be found in~\citep{coogan2016traffic}. 
In this example, the capacity of links 0,1 and 2 are set to 40 and the capacity of links 3,4 and 5 are set to 20, i.e., $x^i \in [0,40] $ for $i \in \{0,1,2\}$ and $x^i \in [0,20] $ $for$ $i \in \{3,4,5\}$. The signals can be 0 or 1, where $s^i = 0$ and $s^i = 1 $ means that traffic flow is allowed in horizontal direction and vertical direction, respectively. Here, our goal is to find reasons which lead to congestion on link $x^1$. For this purpose, we generate a dataset $\mathcal{D}$ as follows. We simulate the system 20 times from random initial conditions for $100$ steps. During the simulations, at each time step the signal values are generated randomly. The traces are labeled according to the following rule:
\begin{align}\label{eq:label_traffic}
	l_t = &\begin{cases}  1 \text{ if } (x_t^1 > 30) \\
	 0 \text{ otherwise}
	\end{cases}
\end{align}
Specifically, the time points at which link 1 has more than 30 vehicles, that is $75\%$ of its capacity, are labeled with 1.
The resulting dataset has $2000$ data points, $456$ of which has label 1.  A sample trace of the system is shown in Fig.~\ref{fig:traf_tra}

We generate the set of all parametric ptSTL formulas $\mathcal{F}^{\leq 2}$ with  at most $2$ parameters over the system variables $\{x^i \mid i=0,\ldots,5\} \cup \{s^0, s^1\}$ according to~\eqref{eq:formula_space}. The parameter domains are defined as:
$p_a, p_b \in \{i  \mid i=0, \ldots,5   \}$  for $\mathbf{A}_{[p_a, p_b]}$,$\mathbf{P}_{[p_a, p_b]}$,  
$p_x \in \{ 5i + 10 \mid i=0, \ldots,6   \}$ for $x \in \{ 0,1,2\}$, 
$p_x \in \{5i + 5 \mid i=0, \ldots,4   \}$ for $x \in \{ 3,4,5\}$. \\
In this case study, we aim at describing system behaviors that leads to congestion on link 1 before it occurs. For this reason, we use $\mathcal{F}^{\leq 2, s} = \{\mathbf{P}_{[1,1]} \phi \mid \phi \in \mathcal{F}^{\leq 2}\}$ as the set of parametric ptSTL formulas. 
We run Alg.~\ref{alg:synthesize} with the traffic system dataset $\mathcal{D}$, the parametric formula set $\mathcal{F}^{\leq 2,s}$, 
error bound $B=20$, and  formula bound $p=3$.

The optimal formula $\phi$ that the algorithm returns is as follows\footnote{The inequalities over the signals are written as equalities to simplify the presentation. $ s^i > 0 $ is equivalent to $ s^i = 1$ since $s^i \in \{0,1\}$. }:
\begin{align}
&\phi = \phi_1 \vee \phi_2 \vee \phi_3   \\  \nonumber
&\phi_1 = \mathbf{P}_{[1,1]}(( x^1 > 15 ) \wedge  ( s^1 = 1 )\wedge  ( s^0 = 0 ))\\  \nonumber 
&\phi_2 = \mathbf{P}_{[1,1]}(( x^1 > 25 ) \wedge  ( s^1 = 1 ))\\  \nonumber 
&\phi_3 = \mathbf{P}_{[1,1]}( ( x^4 <  10 ) \wedge ( s^1 = 1 )  \wedge  ( s^0 = 0 ))\\  \nonumber 
 \end{align}

  Each sub-formula $\phi_1, \phi_2 $ and $ \phi_3$ explains a condition that leads to congestion in link 1 in the current step.  The formulas state that, link 1 will be congested at the next time step ($\phi_1$): if $s^1$ blocks link 1 while  $s^0$ allows flow of vehicles from link 0 to link 1 when there are more than 15 vehicles on link 1, ($\phi_2$):  if  $s^1$ blocks link 1 when there are more than 25 vehicles on it, ($\phi_3$): if $s^1$ blocks link 1 while  $s^0$ allows flow of vehicles from link 0 to link 1 when there are less than 10 vehicles on link 0.

 $TP^{\#}(\phi,\mathcal{D}) $ and $FP^{\#}(\phi,\mathcal{D})$  valuations of formula $\phi$ are 454 and 30, respectively. Total mismatch count of 2000 points is computed as 32.
This whole process took 1205 seconds on the same machine as Ex.~\ref{ex:synthesis}.

\section{Conclusion}\label{sec:conclusion}

In this paper, we developed a framework to synthesize ptSTL formulas from a labeled dataset. We first defined monotonicity properties of the success measures, namely the number of correctly identified positive labels. We developed an efficient algorithm based on monotonicity fo find parameter valuations of a parametric ptSTL formula. Finally, we presented an efficient formula synthesis algorithm that performs parameter synthesis for each parametric ptSTL formula, and iteratively generates a formula as a disjunction of the optimized formulas by considering the evaluation of the combined formula. We showed on two examples that the proposed framework is able to generate complex and long formulas with high accuracy, that was not possible in~\citep{codit2018}.   A future research direction is adapting this method to the datasets in which the exact violation time is unknown.

\end{document}